# Sub-barrier quantum tunneling: eliminating the MacColl-Hartman paradox

A. Zh. Muradyan

Laboratory for Research and Modeling of Quantum Phenomena, Yerevan State University,
1 Alex Manoogian, Yerevan, 0025, Republic of Armenia

E-mail: muradyan@ysu.am

I show that the MacColl-Hartman effect, namely, the saturation of the group delay time of sub-barrier quantum tunneling as a function of the barrier width, comes from the saturating behavior of a more fundamental concept - the phase of the stationary wave function. The explanation of saturation is given based on the decomposition of the stationary wave function into the spectrum of wave numbers and formulation of the initial condition for the direction of propagation of the incident matter wave. It is also shown that the saturation plateau in the case of the sub-barrier wave packet considered by MacColl and Hartman actually has a finite length. After the plateau, the sub-barrier tunneling time monotonically increases with increasing width. This applies both to the maximum's time of the wave packet and to the average tunneling time.

## I. INTRODUCTION

Tunneling through a classically forbidden region is a fundamental and intuitively susceptible phenomenon that quantum-wave theory brings to the physical world. Soon after the discovery, attention was also drawn to the temporal evolution of the phenomenon [1]. MacColl discussed the solutions of sub-barrier tunnelling problem qualitatively [2] and stated that "there is no appreciable delay in the transmission of the packet through the barrier." For a long time, this "strange" pattern may have been perceived as a manifestation of the "strange" nature of the tunneling phenomenon itself, especially since it was impossible to test the dynamics of the process experimentally. Later, the technological advances which made semiconductor thin-film devices possible were the first to indicate the need to revise the problem and, above all, to quantify the tunneling event more precisely. Analytical approximation of the stationary phase [3, 4], as well as numerical simulations used by Hartman in [5] (see also [6]), confirmed the result about a finite short tunneling time (called group delay or phase time), which is saturated with distance. It would be irrational to immediately assign this paradoxical result the status of a general regularity - at least because time is not a dynamic variable in standard quantum theory, and the particle is represented as a non-local wave packet [7-9].

Recent advances in attosecond and ultralow temperature technologies have renewed discussions on the problem of quantum tunneling (see, for example, reviews [10-18]). Part of the reason for this, apparently, is the hope to master the single-particle level of conducting the experiment and use the tunneling time data to navigate the space-time interpretation and applications of the wave function.

In this paper, I'll consider the group delay and the status of the MacColl-Hartman paradox in it. Reference to other definitions of tunneling time will be made only as necessary to illustrate the issue under discussion. Outside of this context, I only note that the objections to group delay, such as "We cannot tell where the transmitted wave packet is at $t=0$, and therefore we cannot say that group delay measures the time required for the wave packet to travel from the input to the output" [13], actually attribute to the wave function a complete analogy with the pulse of the electromagnetic wave and, therefore, a classical interpretation.

## II. GROUP DELAY

Let's start the discussion with a well known procedure of how the delay time is introduced into the theory of quantum tunneling. The system under discussion is an incident wave packet located around some value of coordinate $x_0 < 0$ and a classically forbidden rectangular potential barrier extending from $x=0$ to $x=l$. The wave function $\psi(x,t)$ of the wave packet decomposes into stationary states $\psi_\varepsilon(x)$:

$$\psi(x,t) = \int_0^\infty f(\varepsilon)\psi_\varepsilon(x)e^{i\varepsilon t}d\varepsilon , \qquad (1)$$

where $f(\varepsilon)$ is peaked around some value of energy $E_0 > 0$, $x$, $t$, and $\varepsilon$ are the dimensionless coordinate, time, and energy, normalized to an arbitrary length $L$, the inverse of the "recoil" frequency $\omega_r = \hbar L^2/2m$ (with $m$ the mass of the particle) and the recoil energy $\varepsilon_r = \hbar \omega_r$, respectively. $\psi_\varepsilon(x)$ is assumed to be normalized to the Dirac delta function:

$$\int_0^\infty \psi_{\varepsilon'}^*(x)\psi_\varepsilon(x)dx = \delta(\varepsilon' - \varepsilon). \qquad (2)$$

The wave function $\psi(x,t)$ is normalized to one. And finally, the distribution function $f(\varepsilon)$ is determined by the initial form of the wave packet $\psi(x,0)$:

$$f(\varepsilon) = \int_{-\infty}^{0} \psi_\varepsilon^*(x)\psi(x,0)dx, \tag{3}$$

where it is taken into account that the initial wave packet is located at $x \leq 0$, to the left of the potential barrier. This integral can, in principle, be evaluated once $\psi(x,0)$ is specified.

The wave packet passed through the potential barrier in accordance with the expression (1) is described by a package of stationary states of the form

$$\psi_\varepsilon(x \geq l) = |T(\varepsilon)|\exp\left(i\alpha(\varepsilon) + i\sqrt{\varepsilon}\,x - i\varepsilon t\right), \tag{4}$$

where $|T(\varepsilon)|$ and $\alpha(\varepsilon)$ are, respectively, the probability amplitude and the phase shift of the transmitted wave relative to the incident wave. It is a positive function $|T(\varepsilon)|$ multiplied by the sinusoidal function. The only component $\varepsilon_0$ that makes a significant contribution to the integral (1) is the component corresponding to the maximum of $f(\varepsilon)|T(\varepsilon)|$, for which the phase of the sinusoidal function $\alpha(\varepsilon) + \sqrt{\varepsilon}\,x - \varepsilon t$ is constant. Therefore, the peak coordinate $x_p(t) \geq l$ of the transmitted wave packet in the stationary phase approximation is given by

$$\frac{d\alpha(\varepsilon)}{d\varepsilon} + \frac{1}{2\sqrt{\varepsilon}} x_p - t = 0, \tag{5}$$

where, for brevity, the index zero of the energy $\varepsilon$ is omitted. In the absence of a potential barrier, $T = 1$, $\alpha = 0$, and the remaining equation from (5) describes the free propagation of the wave packet peak with the dimensionless group velocity $v_g = 2\sqrt{\varepsilon}$. This justifies the interpretation of $d\alpha(\varepsilon)/d\varepsilon$ as a temporal delay that the particle acquires from its interaction with the barrier. For reasons that will become clear later, I prefer to use only the term "group delay" of the two accepted terms for the time of movement from the left-hand point of the barrier $x_{p,in} = 0$ to its right-hand point $x_{p,out} = l$. From Eq. (5) it reads

$$\tau_g = \frac{l}{2\sqrt{\varepsilon}} + \frac{d\alpha(\varepsilon)}{d\varepsilon}, \tag{6}$$

It is this time, according to McCall and Hartman, that is saturated depending on the width of the potential barrier $l$ when the particle is tunneling under the barrier. The pattern is illustrated in Fig. 1, comparing it with the free propagation of the same initial wave packet.

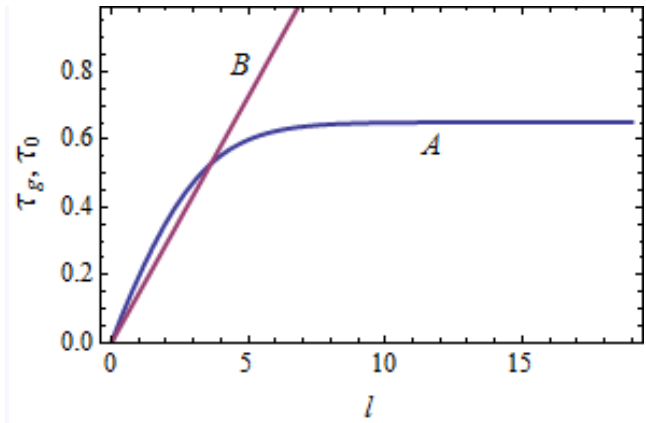

FIG. 1. Group delay $\tau_g$ (A) and free propagation time $\tau_0$ (B) of a wave packet, as a function of rectangular potential barrier dimensionless width $l$. The height of the potential barrier $u_0 = 12$ (in units of recoil energy $\varepsilon_r$), and the energy value $\varepsilon = 11.8$ is slightly lower than this height.

Another well-known regularity is also visible: with a relatively narrow barrier the propagation time is longer than with free propagation. The barrier, as in the classically allowed case, slows down the movement of the particle through the potential region. However, with increasing thickness, especially with asymptotically thick barriers, the behavior becomes, to put it mildly, counterintuitive.

A similar behavior for the sub-barrier group delay (in fact, the group advance relative to free propagation) also holds in the dependence on the energy of the incident particles (Fig. 2).

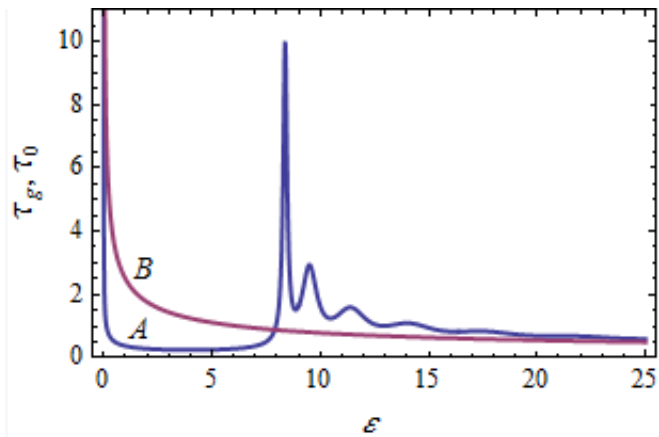

FIG. 2. Group delay $\tau_g$ (A) and free propagation time $\tau_0$ (B) of a wave packet, as function of (dimensionless) energy $\varepsilon$ of the particles incident on the potential barrier. Its height $u_0 = 8$. The intersection point of the graphs, that is, the start of the delay mode relative to the free propagation, is just below the value of $u_0$, at $\varepsilon \approx 7.9$.

It should be noted that the saturating picture for asymptotically expanding barriers also holds for the dwell time $\tau_d$ [19-22], which is defined in the context of 1D elastic scattering [23] as the ratio of the probability that a particle is located in a certain region of space (in the case under consideration, it is classically impenetrable) to the probability current $j_{in}$ incident on this region:

$$\tau_d = j_{in}^{-1} \int_0^l \psi_\varepsilon^*(x)\psi_\varepsilon(x)dx. \qquad (7)$$

This definition assumes that the stream of incident particles is divided in some proportion between the two channels - transmission and reflection. But this is not a precision quantum approach, but a semi-classical view of the tunneling process, since the probability currents are defined for separate, incident, transmitted, and reflected, parts of the wave function. Moreover, this is not the only possible mapping of the group delay definition [23] to the stationary tunneling process. For the transmission problem, a more direct mapping of definition [23] would be expression (7), in which the current $j_{in}$ is replaced by the tunnelling current $j_{out}$ or, equivalently, by the full current $j$:

$$\tau_d = j^{-1} \int_0^l \psi_\varepsilon^*(x)\psi_\varepsilon(x)dx. \qquad (8)$$

Note, by the way, that such a proposal was already made in literature, but it is referred to as a proposal: an ansatz [24, 25], or transmitted flux delay [13]. The time set in this way does not saturate with barrier width.

## III. ELIMINATING THE MACCOLL-HARTMAN PARADOX

Another approach to determining the sub-barrier tunneling time is the introduction of an additional physical quantity into the particle-potential barrier system, the time change of which works as an "internal clock" showing the time spent by the particle inside the potential barrier. Examples are the Larmor rotation of the spin in an applied magnetic field [17-20] and the harmonic generation process in a periodically changing weak field superimposed on the tunneling potential [26, 27, 7]. By contrast, these ones, as well as a number of relevant definitions of tunneling time, yield a result proportional to the barrier width (in the opaque rectangular-barrier limit). That is, there is no MacColl-Hartman paradox for them. Hence, the group delay seems to be unique in the family of tunneling time definitions, and thereby needs to be considered in more detail.

The first question that arises is whether the definition of group delay is an independent definition or is a consequence of a more fundamental concept, namely, the running phase of a monochromatic wave. To figure this out, we turn to the stationary wave functions of the particle $\psi_\varepsilon(x,t) = \psi_\varepsilon(x)e^{-i\varepsilon t}$ and follow the development of its phase. To the right of the barrier (including the endpoint of interest $z = l$), the wave function has been written in the form (4). Then to the left of the barrier, the incident wave should be presented by the expression $\exp(i\sqrt{\varepsilon}\,x - i\varepsilon t)$, where the starting point of time coincides with the choice in the right-hand form of (4). This is firstly because the phase $\alpha(\varepsilon)$ is solely due to the interaction of the particle with the potential barrier; and secondly, because the wave reflected from the barrier is not reflected further and therefore cannot have any contribution to the tunneled wave (4). The inclusion of the measurement process in the scope of the issue under discussion leaves this statement unchanged.

Let's denote $t_{in}$ - the moment when a certain wave front reaches the beginning of the barrier $x_{in} = 0$, and $t_{out}$ - the moment when the same wave front reaches the end of the barrier $x_{out} = l$. Then we have the equality $-i\varepsilon t_{in} = i\alpha(\varepsilon) + i\sqrt{\varepsilon}\,x_{out} - i\varepsilon t_{out}$ and, accordingly, the expression

$$t_{ph} = \frac{\alpha(\varepsilon)}{\varepsilon} + \frac{l}{\sqrt{\varepsilon}} \qquad (9)$$

for the tunneling time $t_{ph} = t_{out} - t_{in}$. I think, it makes sense to call it "phase time". The phase shift term $\alpha(\varepsilon)$ in our notations is given by

$$\alpha(\varepsilon) = \frac{4i\,e^{(-ik+\chi)l}\,k\,\chi}{-(k-i\chi)^2 + e^{2\chi l}(k+i\chi)^2}, \qquad (10)$$

where $k = \sqrt{\varepsilon}$, $\chi = \sqrt{u_0 - \varepsilon}$, and $\varepsilon < u_0$. The dependence of $t_{ph}$ on the width of the barrier is illustrated in Fig. 3, graph A (together with graph B of free propagation). It is being saturated!

Now let us pay attention to the fact that the condition (6) defining the group delay is, at least on a formal mathematical basis, the energy derivative of the condition for determining the phase time (9). Hence, it can be argued that the saturable character of $t_g$ is a direct consequence of the saturable property of $t_{ph}$, and the question of saturation turns from the original subject of group delay into a simpler subject of phase time. This means that the answer to the paradoxical behavior of the time of sub-barrier tunneling can be sought in phase propagation, which is easier to study. In this regard, we first note that the general solution of the stationary Schrodinger equation of the barrier region, being a sum of two real exponential functions, can be decomposed in the $k$-space of wave vectors and be represented as a superposition of two groups of spectral components of oppositely propagating waves: passing through the barrier and reflecting from it. Since the wave incident on the barrier propagates from left to right, the spectrum of positive wave vectors will be more powerful than the spectrum of negative wave vectors. This picture of a stationary state makes it possible to determine its time of sub-barrier propagation, proceeding, for example, from the expression of the zero phase: $kx - \varepsilon t = 0$ for any $k$ of the spectrum. For waves propagating to the right $k > 0$, and as a result, a positive value $t = x/k$ is obtained for the propagation time. All components of the spectrum of wave vectors passing the barrier from left to right give a positive contribution to the $t_{ph}$ value. For opposite components, propagating from right to left, we must take $k < 0$. Then, under the above zero phase condition (defined for the case of left-to-right propagation), $t = -x/|k|$. The contribution of right-to-left components to the phase time $t_{ph}$ is negative. Since the proportion of reflected components gradually increases with increasing barrier width, their negative contribution also gradually increases, leading the total contribution to some finite asymptote in the limit of wide barriers. Finally, since the $k > 0$-space is more powerful than the $k < 0$-space, the asymptotic value of the tunneling time is positive.

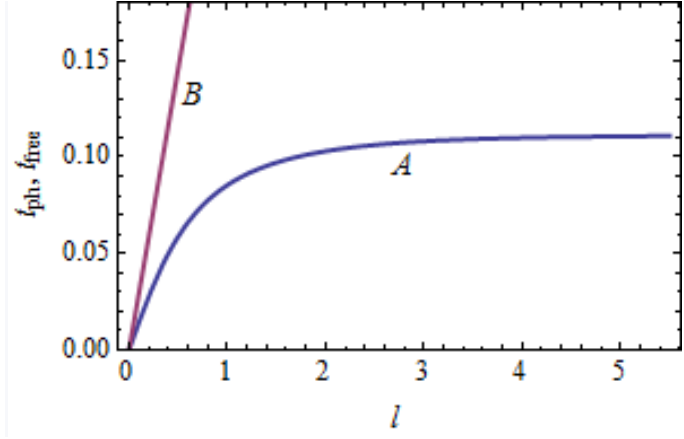

FIG. 3. Phase time $t_{ph}$ in the presence of a potential barrier ($A$) and free propagation phase time $t_{free}$ ($B$) as a function of the barrier width $l$. The regularity of saturation holds for any value of the energy $\varepsilon$, as long as the condition of sub-barrier propagation $\varepsilon < u_0$ is satisfied. The parameters are the same as in Fig. 1.

Thus, we can conclude that the MacColl-Hartman paradox is a consequence of the stationary phase approximation, which leads the wave packet propagation problem to the stationary state propagation problem. In turn, the saturating character of the time of sub-barrier tunneling, depending on the width of the barrier, is simply explained by the opposite contribution of direct and reverse traveling waves. This is strictly from a mathematical point of view, but it contradicts classical intuition.

As a logical continuation, we can state that the tunneling time of the wave packet $t_{max}$ should repeat the saturating course of the stationary states at least up to some significant width of the barrier. But after that, the tunneling time can, in principle, increase, thereby eliminating the paradoxical behavior discovered by MacColl and Hartman. This would be the most natural way out of the current situation.

To check the validity of this expectation, I performed numerical calculations using the standard mathematical package Wolfram Mathematica. The obtained dependence on the barrier width, shown in Fig. 4, has exactly the expected form. Namely, the sub-barrier tunneling time has a plateau depending on the width of the potential barrier and the plateau has a finite length. and at that, the subsequent increase in the barrier passage time occurs at an accelerating rate. It should also be noted that the subsequent growth of the tunneling time runs at an accelerating rate.

Somewhat unexpectedly, this accelerating character strongly resembles the result of [28] obtained in Bohm's interpretation of quantum mechanics. Consistency aspects will be considered in a separate study.

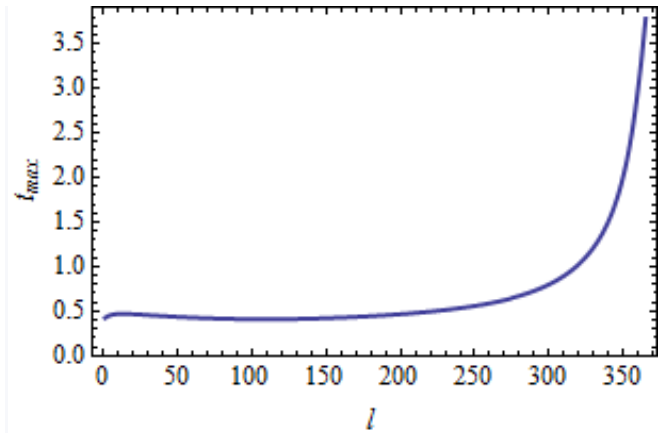

FIG. 4. The tunneling time of the maximum of the wave packet, obtained by numerical calculations based on the general formulas (1)-(3). The wave packet at the initial moment of time is directly adjacent to the potential barrier and has the form of a sinusoid at half of its period: $\psi(x,0) = A[1-\cos(2x/b)]e^{ipx}$, $-\pi b < x < 0$, where $p$ is the average momentum of the wave packet before the collision, $A$ and $b$ are the amplitude and halfwidth, respectively. The graph refers only to sub-barrier tunneling, that is, the upper limit of the integral in the formula (1) from infinity is replaced by $u_0$. The values of the system parameters $u_0 = 31.4$, $p = 3.6$, and $b = 2$ are chosen so that the initial shape of the sub-barrier wave packet practically does not differ from the exact shape of (1). For ease of viewing, the time of passage of the sub-barrier $t_{ph}$ is presented in sum with approximately constant time $t_{in}$ of the arrival of the maximum of the wave packet from the initial position to the beginning of the barrier.

The regularity for the tunneling time of the maximum of a sub-barrier wave packet (exact or in the approximation of the stationary phase) directly implies the question of the average tunneling time. I define the average time of passage of a given point $x$ by a wave packet (1) as the quantum mechanical average of the moment of time $t$ (despite the absence of the dynamic status of the latter in the formalism of quantum mechanics):

$$t_{mean}(x_0; x) = \frac{\int_0^\infty t \, |\psi(x,t)|^2 \, dt}{\int_0^\infty |\psi(x,t)|^2 \, dt}, \qquad (11)$$

where $x_0 < 0$ is some starting point, the average position of the wave packet at initial time t=0. The time (11) with $x = l$ is the analog of the time, pictured in Fig. 4 and is illustrated depending on the width of the potential barrier in Fig. 5. (In fact, to get the time of sub-barrier tunneling, one should subtract from the present graph its initial value).

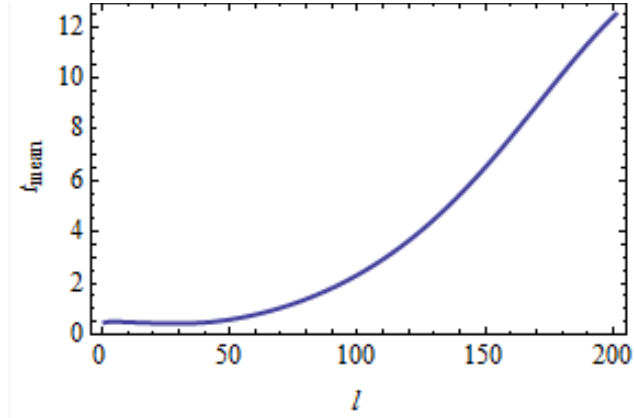

FIG. 5. Average time of passage of the endpoint of a potential barrier by a sub-barrier wave packet. The parameters are the same as in Fig. 4.

As is seen, the average time of sub-barrier propagation also demonstrates, depending on the thickness of the barrier, an initial plateau followed by a monotonous growth. It can also be noted that after the plateau this time increases noticeably faster than the time in Fig. 4. This is due to a number of noticeable sub-packets following the main one, pronounced wave packet at the end of the potential barrier, the weight and length of which increase with increasing barrier width.

## IV. SUMMARY

It is shown that the plateau-shaped regularity, discovered by MacColl and Hartman for the time of sub-barrier tunneling in dependence of the potential barrier width, is the prerogative of not the propagation of the maximum of the wave packet, but of a more fundamental concept: the phase of a stationary wave function. Then a physical explanation of the formation of the plateau is given, based on the Fourier decomposition of the stationary wave function of the barrier region in the space of wave numbers and on the fact that the boundary condition of the phase evolution is given for one-way motion, from left to right. Then the wave numbers of direct propagation give positive contributions to the time of propagation, while the wave numbers of the reflected direction give negative contributions and thereby reduce the total time of sub-barrier propagation. So, the interference of matter waves in case of classically forbidden motion possibility of reflection behaves completely opposite to classical concepts, namely, instead of lengthening, it reduces the propagation time. I call it counterintuitive interference. The distribution of the wave numbers gradually approaches a certain shape with the increase in the width of the barrier, and the time of the sub-barrier propagation, being the result of counterintuitive, also slows down in growth and goes to a definite plateau.

When we are dealing with a realistic space-time wave packet, same as the grouped distribution of stationary states, the plateau for the passage time of the maximum of the wave packet, as well as for the average quantum mechanical time, has a finite length. After the plateau, the time of sub-barrier tunneling monotonically increases with increasing barrier width. This can be interpreted on the base of stationary solutions as follows. For each of them, the spectrum of negative wave numbers gives a negative contribution into the duration of tunneling, but their share for wide barriers is saturated, thereby limiting the negative contribution. The slowly narrowing spectrum of positive wave numbers continues to form a wave packet moving to the right.

## ACKNOWLEDGMENTS

This work was financially supported by the Science Committee of the Republic of Armenia

within the framework of the Laboratory for Research and Modeling of the Quantum Phenomena at Yerevan State University.

________________________________________